\documentclass[prb,twocolumn,showpacs,floatfix]{revtex4}%
\usepackage{graphicx}
\usepackage{dcolumn}
\usepackage{bm}

\begin{document}

\title{Spectral densities of Kondo impurities in nanoscopic systems}
\author{P. S. Cornaglia}
\author{C. A. Balseiro}

\affiliation{Instituto Balseiro and Centro At\'omico Bariloche, Comisi\'on
Nacional de Energia At\'omica, 8400 San Carlos de Bariloche, Argentina.}

\date{\today}

\begin{abstract}
We present results for the spectral properties of Kondo impurities in nanoscopic systems. Using Wilson's renormalization group we analyze the frequency and temperature dependence of the impurity spectral density $\rho _{i}(\omega ,T)$ for impurities in small systems that are either isolated or in contact with a reservoir.  We have performed a detailed analysis of the structure of $\rho _{i}(\omega ,T)$ for $\omega $ around the Fermi energy for different Fermi energies relative to the intrinsic structure of the local density of states. We show how the electron confinement energy scales introduce new features in the frequency and temperature dependence of the impurity spectral properties.

\end{abstract}


\maketitle

\section{Introduction}

The physics at the nanoscale has emerged as one of the most active areas in
condensed matter physics. This new field includes the study of small
metallic and superconducting islands, quantum dots and
nanostructurated semiconductors, nanotubes and quantum corrals, nanoelectronics and nano-characterization among other things.\cite{nanoscience}

The advances in nanotechnologies revived the interest in the Kondo effect,%
\cite{Hewson1993} one of the paradigms of strongly correlated systems. On one
hand, Scanning Tunneling Microscopy (STM) allowed the direct measurement of
local spectroscopic properties of Kondo impurities on noble metal surfaces%
\cite{madhavan98,nagaoka2002} and in nanoscopic systems.\cite{Manoharan2000} On the other hand, it has
been shown that single electron transistors and single walled carbon
nanotubes weakly coupled to contacts may behave as Kondo impurities
generating new alternatives to study the phenomena.\cite{Meir1991}

In its simplest form, the Kondo problem is a single quantum spin interacting
with an ideal electron gas. An antiferromagnetic coupling between the
impurity and the free electron spins gives rise to an anomalous scattering at the
Fermi energy leading to a large impurity contribution to the
resistivity. Simultaneously the impurity spin is screened by the conduction
electrons and the magnetic susceptibility saturates at low temperature.
There is a characteristic temperature $T_{K}$ that separates the low
temperature from the high temperature regimes. At $T\gg T_{K}$, the
impurity spin is essentially free and the problem can be treated by
perturbations in a dimensionless coupling constant $\lambda $. At $T\ll T_{K}$
the impurity spin is screened forming a singlet complex with the conduction
electrons and the system is described by an infinite effective coupling. The
crossover regime with $T\sim T_{K}$ is more difficult to describe and the
best treatment corresponds to the numerical renormalization group (NRG) as
done by Wilson.\cite{Wilson1975}

The characteristic Kondo temperature is given by $T_{K}$ $\sim D\sqrt{%
\lambda }e^{-1/\lambda }$ where $2D$ is the free electron bandwidth.
Associated to this energy scale there is a characteristic length scale known
as the Kondo screening length $\xi _{K}=\hbar v_{F}/$ $k_{B}T_{K}$ where $%
v_{F}$ is the Fermi velocity. The physical meaning of the screening length
is that, in the low temperature regime where the impurity spin is screened,
the antiferromagnetic correlations between the impurity and conduction
electron spins extend up to a distance of the order of $\xi _{K}$.

The problem of Kondo impurities in nanoscopic systems for different
experimental realizations has been the subject of many recent theoretical
and experimental works.\cite{Thimm1999} When a Kondo system, either an atomic impurity or an
artificial atom like a quantum dot (QD), is embedded in a small system of volume $L^{d}$
where $d$ is the spacial dimension, the length-scale $L$ should be
compared with the characteristic Kondo length $\xi _{K}$. For $L<\xi _{K}$
finite-size effects are expected to be important. The condition $L\sim \xi
_{K}$ is equivalent to $k_{B}T_{K}\sim \Delta $ where the energy $\Delta $
gives the average level spacing of the finite system. For a finite system
the characteristic energy $\Delta $ acts as a low energy cutoff for the
charge and spin excitations and consequently it modifies the low temperature
behavior of the system.

The ground state properties of a Kondo impurity in a small system have been
addressed by a number of authors using different approximations.\cite{Thimm1999, Cornaglia2002, Affleck2001} The
thermodynamic properties and the effect of coupling the system to a
macroscopic reservoir have recently been studied using the Wilson's
renormalization group.\cite{Cornaglia2002} In this work we extend renormalization group
calculations to evaluate the impurity spectral density and analyze how the
new energy or length scales introduced by the finite size affects the Kondo
resonance at the Fermi energy. We also study the temperature dependence of
the low energy spectral density.

The rest of the paper is organized as follows: in section II we present the
model and describe how the numerical renormalization group is adapted to our
case. We then recap the most relevant thermodynamic properties.
Section III contains the spectral properties of the impurity for frequencies close to the
Fermi energy. After presenting the general formulation we show results for
finite systems, systems in contact with a reservoir and the temperature
dependence of the Kondo resonances. Finally section IV includes a summary
and discussion.

\section{Model and Thermodynamic properties}

In this section we present the model for an impurity in a nanoscopic system
coupled with a macroscopic reservoir and briefly discuss how Wilson's
renormalization group is adapted to this problem. Then we review the
thermodynamic properties of the system.

\subsection{Model Hamiltonian and Wilson's Renormalization Group}

Our starting point is the Anderson model for magnetic impurities with a
Hamiltonian which in the usual notation reads:\cite{Anderson1961}

\begin{eqnarray}\label{hamil}
H_{AM} &=&\sum_{\sigma }\varepsilon _{d}d_{\sigma }^{\dagger }d_{\sigma
}+Ud_{\uparrow }^{\dagger }d_{\uparrow }d_{\downarrow }^{\dagger
}d_{\downarrow }+\sum_{\nu ,\sigma }\varepsilon _{\nu }c_{\nu \sigma
}^{\dagger }c_{\nu \sigma }  \nonumber \\
&&+\sum_{\nu ,\sigma }(V_{\nu }^{*}c_{\nu \sigma }^{\dagger }d_{\sigma
}+V_{\nu }d_{\sigma }^{\dagger }c_{\nu \sigma })-\mu _{i}BS_{iz},
\label{ham}
\end{eqnarray}
where the operator $d_{\sigma }^{\dagger }$ creates an electron with spin $%
\sigma $ at the impurity orbital with energy $\varepsilon _{d}$ and Coulomb
repulsion $U$, and $c_{\nu \sigma }^{\dagger }$ creates an electron in an
extended state with quantum numbers $\nu $ and $\sigma $ and energy $%
\varepsilon _{\nu }$. The last term represents the effect of an external
magnetic field along the $z$-direction coupled to the impurity spin ${\bf S}%
_{i}$. Hereafter we will use $D=1$ as our unit of energy.

\begin{figure}[tbp]
\includegraphics[width=7.5cm,clip=true]{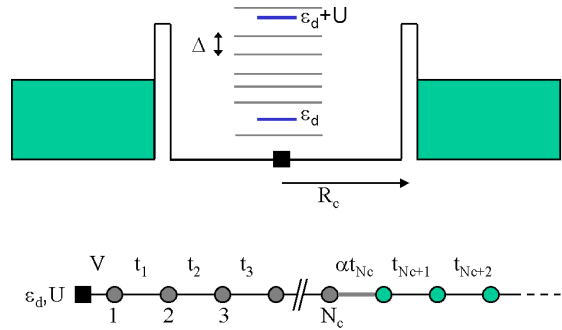}
\caption{ A sketch of the central grain embedded in a metal (a).
Inside the grain the characteristic energy level spacing $\Delta $ is shown
together with the impurity levels at $\varepsilon _{d}$ and $\varepsilon
_{d}+U$. In (b) the linear chain obtained after Wilson's canonical
transformation.}
\label{fig1}
\end{figure}

In this notation, the nanostructure of the system is hidden in the
structure of the one-electron extended states with energies $\varepsilon
_{\nu }$ and wavefunctions $\psi _{\nu }({\bf r})$. In equation (\ref{ham})
the hybridization matrix elements are taken proportional to the extended
state wavefunctions at the impurity position,  i.e. $V_{\nu }=V_{0}\psi
_{\nu }(0)$ where the impurity position is defined as the origin of
coordinates. We consider a simple structure consisting of a spherical
metallic cluster of radius $R_{c}$ with the impurity at the center. The
cluster is embedded in a bulk material with which it is weakly coupled
through a large surface barrier. The Hamiltonian can be put in the form $%
H_{AM}=H_{AM}^{0}+W(r,R_{c})$ where the first term is the Anderson model
Hamiltonian for an impurity in an infinite homogeneous host and the last
term is a spherically symmetric potential barrier placed at a distance $R_{c}
$ from the impurity. Figure \ref{fig1}(a) illustrates the configuration described by
the model. For an infinite impenetrable barrier, the central cluster is
decoupled from the macroscopic reservoir and the model describes an impurity
in a small sample. In this situation, the extended states that are coupled to the impurity, are confined in
the central cluster and their energy spectrum is a discrete spectrum with a
mean energy level separation given by the characteristic energy $\Delta .$
For a finite barrier these states are hybridized with the continuum acquiring a finite lifetime, the local density of states inside the
central cluster then presents resonances separated by the energy $\Delta $ and
with widths $\gamma $ that are determined by the barrier. The model then
incorporates the new energy scales $\Delta$ and $\gamma$, that may drastically change the impurity behavior.

This geometrical structure is particularly appropriate to use the numerical
renormalization group approach developed by Wilson. Wilson introduced a logarithmic discretization of the energy of the conduction electrons, dividing the band in a series of energy intervals with exponentially decreasing width.
By means of a canonical transformation the resulting Hamiltonian can be mapped into a linear chain with variable hoppings. Each site representing an orbital surrounding the impurity with an associated energy (or length) scale. 
Wilson proposed to solve the linear chain by iterative perturbation. A truncated chain with $N$ sites, described by an effective Hamiltonian $H_{N}$, gives the correct physics on an energy scale $\tilde{\omega} _{N}$. A renormalization group
transformation  corresponds to adding a site to the chain and relates the Hamiltonians describing successive
lower energy scales. This leads to a systematic way
of calculating the thermodynamic properties at successive lower temperatures
and the spectroscopic properties at successive lower frequencies.

In our case, the Hamiltonian $H_{AM}^{0}$ is rewritten in Wilson's basis as
schematically shown in Fig.\ref{fig1}(b). The potential barrier is described
including a higher diagonal energy to the orbital centered around $R_{c}$.
Due to the structure of Wilson's basis wavefunctions, a given potential barrier
leads to a diagonal energy that decreases as $R_{c}$ increases.
Alternatively the barrier can be simulated with a smaller hopping matrix
element in the region where $W(r,R_{c})$ is different from zero. We adopted
this last description reducing one hopping term by a factor $\alpha $. In what follows it is assumed that the band of extended states
is half filled - the Fermi energy is set equal to zero - and $\epsilon_d=-U/2$. This guarantees that the electron-hole symmetry of the problem is preserved.

\begin{figure}[tbp]
\includegraphics[width=7cm,clip=true]{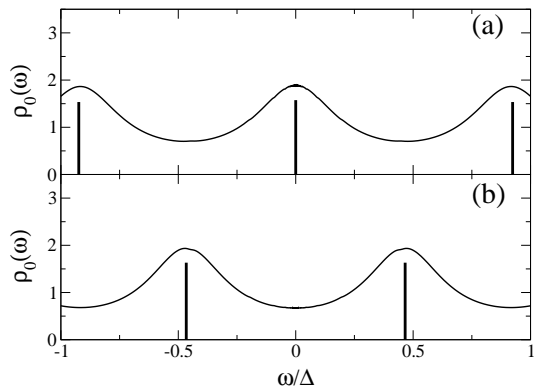}
\caption{Unperturbed local density of extended states at the impurity coordinate
(site 1) for a cluster with $\alpha =0$ (vertical bars) and $\alpha=0.05$ (lines). In the later case the spectral density has been evaluated using a small imaginary part in the frequency (a) $N_{c}=19$ and (b) $N_{c}=20$.}
\label{fig2}
\end{figure}

The properties of a Kondo impurity depend on the local density of states at
the impurity coordinate that in Wilson's representation is the local density
of states at site 1. We end this section with a brief discussion on the
effect of the confining potential $W(r,R_{c})$ on the local density of
states close to the Fermi energy. For a given $N$ the
spectrum of $H_{N}$ depends on the parity of $N$. In the absence of impurity,
a one particle state at zero energy exists only for odd $N$ as schematically shown in
Fig. \ref{fig2}. The spectrum, as a function of $N$, alternates between those of
Figs. \ref{fig2}(a) and \ref{fig2}(b). The mean energy separation, between the one-electron states, is given by the characteristic energy $\Delta $ that decreases as $N$ increases. This is the
spectrum of a finite system ($N = N_c$) described by an infinite barrier. The other one electron states, not shown in Fig. \ref{fig2}, are not at energies $n\Delta$ or $(n+1/2)\Delta$ with integer $n$. This is due to the logarithmic discretization of the band. How the NRG can be used to analyze finite size effects was shown by Nozieres\cite{Nozieres}. To show that the logarithmic discretization  correctly describes the low energy spectrum of a impurity in finite systems, in the next section, we compare the exacts results on a finite system with those obtained with the NRG. The alternating spectrum with a fixed Fermi energy corresponds
to an alternating parity in the number of electrons, in fact for a fixed
electron density, as the radius $R_{c}$ of the cluster increases, the number
of electrons alternates between even and odd. We recall that due to the
symmetry of the system only the sector of the Hilbert space with s-wave
electrons is considered here.
With a finite barrier ($\alpha \neq 0$) the discrete states of the cluster
are hybridized with the continuum of the host metal and become resonances.
The local density of states then becomes a continuum but retains some
structure characterized by the energy $\Delta $. In the NRG approach,
Hamiltonians $H_{N}$ with $N>N_{c}$ accumulate states at low energies
representing the broadening of the central peak for odd $N_{c}$ or the tails
of the states at $\pm \Delta /2$ for even $N_{c}$. The local densities of
states as obtained with the NRG are also shown in Fig. \ref{fig2}. For $\alpha \neq
0$ the number of electrons in the central cluster is no longer a good
quantum number and in what follow we refer to the situations of Figs. \ref{fig2}(a)
and \ref{fig2}(b) as the Fermi energy being at a resonance (at-resonance) or between
two resonances (off-resonance) respectively.

As we show below, this structure of the local density of states determines
the thermodynamic and the spectral properties of the impurity.

\subsection{Thermodynamic Properties}

Here we briefly recap the thermodynamic properties of the model. As stated
above, for $\alpha =0$ the at-resonance situation corresponds to an odd
number of extended electrons. An impurity in the Kondo limit contributes
with an extra electron and the ground state is a singlet indicated as $%
|0\rangle .$ The expectation value of the impurity spin $\langle
0|S_{iz}|0\rangle $ is zero reflecting the complete screening of the
impurity spin and the zero temperature susceptibility is finite. For the off-resonance situation the ground state of the isolated cluster with a Kondo
impurity is a Kramers spin-$1/2$ doublet, $|\Uparrow \rangle $ and $%
|\Downarrow \rangle $. The expectation value $\langle \Uparrow
|S_{iz}|\Uparrow \rangle =-\langle \Downarrow |S_{iz}|\Downarrow \rangle $
is different from zero and, in the low temperature limit the impurity
susceptibility diverges as $\chi =\mu _{i}^{2}\langle \Uparrow
|S_{iz}|\Uparrow \rangle ^{2}/k_{B}T$.

For a finite barrier the susceptibility always saturates, however the at-resonance and off-resonance situations give rise to very different
temperature dependences. We calculated the impurity magnetic susceptibility
given by \cite{Desc} 
\begin{eqnarray}
k_{B}T\chi  &=&{\mu }_{i}^{2}(\sum_{\nu }P_{\nu }|\langle \nu |S_{iz}|\nu
\rangle |^{2}+  \nonumber \\
&&2k_{B}T\sum_{\nu \neq \xi }|\langle \nu |S_{iz}|\xi \rangle |^{2}\frac{%
P_{\nu }}{E_{\xi }-E_{\nu }}),  \label{susc}
\end{eqnarray}
where the summation is done over the low energy states $|\nu \rangle $ with
energies $E_{\nu }$ and $P_{\nu }=\exp (-E_{\nu }/k_{B}T)/Z$. The matrix
elements $\langle \nu |S_{iz}|\xi \rangle $ have to be evaluated in a
recursive way at each renormalization step. The susceptibility reflects the
thermodynamic properties of the system and we use the effective magnetic
moment ${\mu }^{2}=k_{B}T\chi $ as an indicator of the degree of screening
of the impurity spin.

\begin{figure}[tbp]
\includegraphics[width=7cm,clip=true]{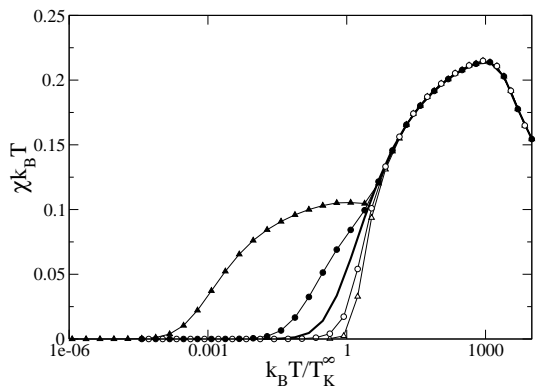}
\caption{ The magnetic susceptibility $k_{B}T\chi $ versus temperature. The
thick line corresponds to an impurity in a bulk material ($\alpha =1$),
lines and symbols to a barrier enclosing $N_{c}=19$ and $20$ shells (open
and full symbols respectively). Triangles (circles) correspond to $\alpha
=0.35$ $(0.6)$. The other parameters are: $\varepsilon _{d}=-0.5$, $U=1.0$, and $V=0.2$.}
\label{fig3}
\end{figure}

\begin{figure}[tbp]
\includegraphics[width=7cm,clip=true]{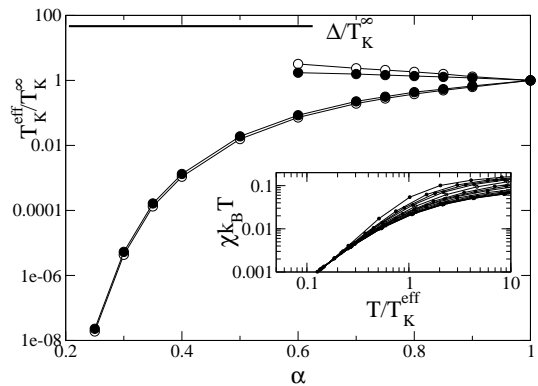}
\caption{The effective Kondo temperature, that describes the low temperature
behavior of the susceptibility, versus the barrier parameter $\alpha $ for
the at-resonance (upper curves) and the off-resonance (lower curves) cases. Open symbols were obtained using Eq. (\ref{teff}), full symbols using the
scaling shown in the inset. The parameters are the same as in Fig. \ref{fig3}.}
\label{fig4}
\end{figure}

In absence of barrier - corresponding to the infinite homogeneous system -
the characteristic energy scale is the Kondo temperature indicated as $%
k_{B}T_{K}^{\infty }$. A finite barrier at $R_{c}$ - enclosing $N_{c}$
shells- introduces the new energy scale $\Delta \cong D\Lambda ^{-N_{c}/2}$, where $\Lambda=2$ is Wilson's discretization parameter.
For $k_{B}T_{K}^{\infty }\gg \Delta $ the fine structure (on the scale of $%
T_{K}^{\infty }$) of the density of states does not change the properties of
the system. Conversely, for $k_{B}T_{K}^{\infty }\sim \Delta $ new
confinement induced regimes are observed: for the system at-resonance, as
the temperature is lowered, there is a rapid decrease in the magnetic moment 
${\mu }^{2}$ when $k_{B}T\sim \Delta $; for the system off-resonance, as $%
k_{B}T$ approaches $\Delta $ the magnetic moment saturates leading to a
plateau in the temperature dependence of ${\mu }^{2}$, only at lower
temperatures the screening is completed (see Fig. \ref{fig3}). This plateau can
be interpreted as the behavior of an isolated cluster. Only at very low
temperatures the coupling with the host states becomes relevant and the
complete screening can occur. The condition for the existence of these new
regimes $k_{B}T_{K}^{\infty }\sim \Delta $ can be put in the form $\xi
_{K}^{\infty }$ $\sim R_{c}$ where $\xi _{K}^{\infty }=\hbar
v_{F}/k_{B}T_{K}^{\infty }$ is the Kondo screening length of the infinite
system. In other words, only if the Kondo screening length is of the order or larger than the system size, the confinement effects become evident in the thermodynamic properties.
Although the behavior of the system is not universal, at very low
temperatures the susceptibility can be used to define an effective Kondo
temperature. In fact the low temperature tail of $k_{B}T\chi $ can be scaled
to define the effective energy scale $T_{K}^{eff}$ for the low energy spin
excitations. We have done this using two different criteria, one is simply
to use Wilson's result:
\begin{equation} \label{teff}
\chi (T=0)=0.103\frac{{\mu }_{i}^{2}}{k_{B}T_{K}},
\end{equation}
to relate the low temperature susceptibility to an effective $T_{K}^{eff}$.
The other alternative is to plot ${\mu }^{2}$ vs. $T/T_{K}^{eff}$ fitting $%
T_{K}^{eff}$ to have a good scaling at low temperatures. The effective Kondo temperatures
describing the low temperature behavior are shown in Fig. \ref{fig4} as a function
of the barrier height for the at-resonance and off-resonance situations. For 
$\alpha >0.2$ the two criteria give the same results, for smaller values of $%
\alpha $ the universal behavior is obtained only at extremely low
temperatures where numerical errors become important. 

\section{Spectral Densities of Impurities in Nanoscopic Systems}

Using the standard definitions and notation, the impurity Green's function
can be written, using the Lehmann representation, as:

\begin{equation}
G_{\sigma }(\omega ,T)=\frac{1}{Z(T)}\sum_{\lambda ,\lambda ^{\prime
}}|\langle \lambda |d_{\sigma }|\lambda ^{\prime }\rangle |^{2}\frac{%
e^{-E_{\lambda }/k_{B}T}+e^{-E_{\lambda ^{\prime }}/k_{B}T}}{\omega
-(E_{\lambda ^{\prime }}-E_{\lambda })},  \label{gi}
\end{equation}
where $Z(T)$ is the grand partition function 
\begin{equation}
Z(T)=\sum_{\lambda }e^{-E_{\lambda }/k_{B}T}.  \label{z}
\end{equation}
The corresponding impurity spectral density is 
\begin{eqnarray}
\rho _{i\sigma }(\omega ,T) &=&-\frac{1}{\pi }[G_{\sigma }(\omega
+i0^{+},T)-G_{\sigma }(\omega -i0^{+},T)]  \nonumber \\
&=&\frac{1}{Z(T)}\sum_{\lambda ,\lambda ^{\prime }}|M_{\lambda ,\lambda
^{\prime }}|^{2}(e^{-E_{\lambda }/k_{B}T}+e^{-E_{\lambda ^{\prime
}}/k_{B}T})\times   \nonumber \\
&&\delta (\omega -(E_{\lambda ^{\prime }}-E_{\lambda })),  \label{ai}
\end{eqnarray}
with $M_{\lambda ,\lambda ^{\prime }}=\langle \lambda |d_{\sigma }|\lambda
^{\prime }\rangle $. In the absence of an external magnetic field, the
impurity spectral density is spin independent and from here on we drop the
spin index in $\rho _{i\sigma }(\omega ,T)$.

We now briefly discuss the application of the NRG to evaluate these
quantities. The zero temperature limit of expression (\ref{ai}) can be put
in the form:
\begin{eqnarray}
\rho _{i}(\omega ,T=0) &=&\frac{1}{Z(0)}\sum_{\lambda ,0}|M_{\lambda
,0}|^{2}\delta (\omega +E_{\lambda })  \nonumber \\
&&+|M_{0,\lambda }|^{2}\delta (\omega -E_{\lambda }),  \label{ai0}
\end{eqnarray}
where the subindex $0$ indicates the ground state, the summation is over all excitations $\lambda $ with energies $E_{\lambda }$ and on
the components of the ground state in the case of a degenerate ground state, 
$Z(0)$ is the zero temperature partition function that gives the ground
state degeneracy. We recall that in the NRG the chemical potential is set at
zero and all many body energies $E_{\lambda }$ are measured from the ground
state energy.

In order to evaluate the impurity spectral density of the real system using
the results of the NRG, consider the spectral density $\rho _{i}^{N}(\omega
,T)$ corresponding to Hamiltonian $H_{N}.$ As mentioned above, the RG
strategy of an iterative diagonalization of a sequence of Hamiltonians $H_{N}
$ with $N=0,1,\ldots$ is based on the fact that, on an energy scale $\tilde{\omega} _{N}
$, the spectrum of $H_{N}$ is representative of the spectrum of the infinite
Hamiltonian $H$. Hence, for $\omega \approx \tilde{\omega} _{N}$ it is a good
approximation to take\cite{Hofstetter} 
\[
\rho _{i}(\omega ,T=0)=\rho _{i}^{N}(\omega ,T=0).
\]
A typical choice is to consider for the $N^{th}$ iteration the energy range $%
\Lambda^{-N/2} <\omega <2\Lambda^{-N/2} $ , that is to consider all the states of $H_{N}$
with energies $E_{\lambda }^{N}$ in this range and all the corresponding
matrix elements $M_{\lambda ,0}^{N}$ to compute the spectral density as 
\begin{eqnarray}
\rho _{i}(\omega ,T &=&0)=\frac{1}{Z_{N}(0)}\sum_{\lambda ,0}|M_{\lambda
,0}^{N}|^{2}\delta (\omega +E_{\lambda }^{N})  \nonumber \\
&&+|M_{0,\lambda }^{N}|^{2}\delta (\omega -E_{\lambda }^{N}).  \label{ai0n}
\end{eqnarray}

Since for the infinite systems the energy spectrum consists of a
continuum, at each energy scale $\tilde{\omega} _{N}$, the discrete spectra are usually
smoothed by replacing the delta function $\delta (\omega -E_{\lambda }^{N})$
by a smooth distribution $P_{N}(\omega -E_{\lambda }^{N})$. At each $N$ we use a Gaussian logarithmic distribution with a width that decreases as the characteristic energy scale of $H_N$. It is
then clear that at high energies, say the atomic energy of the impurity
level $\varepsilon _{d}$, the NRG cannot resolve any detailed structure,
only at low frequencies there is enough resolution to see, for example,
confinement effects. We analyze in detail the impurity spectral density
around the Fermi energy which in fact contains all the physics at an energy
scale relevant for the Kondo effect and for the electron confinement in a
nanoscopic sample.  

Before presenting the renormalization group results we show, as a reference
calculation, some exact results in finite systems.

\subsection{Exact results in finite systems}

Here we present the zero temperature results for an impurity in a finite
system consisting of a linear chain of $N$ equivalent sites with the
impurity at one end. In this case the energy spectrum of extended states $%
\varepsilon _{\nu }$ and the hybridization matrix elements $V_{\nu }$ of
Hamiltonian (1) are given by: 
\[
\varepsilon _{\nu }=-2t\cos (\nu \pi /(N+1)), 
\]
and 
\[
V_{\nu }=V_{0}\sin (\nu \pi /(N+1)),
\]
where $t$ is the hopping matrix element in the chain and $\nu =1,2,....,N$.

\begin{figure}[tbp]
\includegraphics[width=7cm,clip=true]{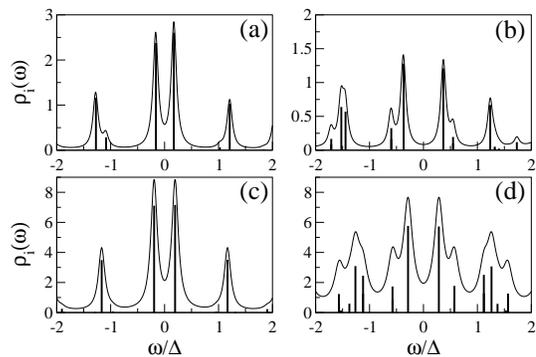}
\caption{Impurity spectral density $\rho _{i}(\omega ,T=0)$. Upper panels
are exact results for a linear chain with $N=8$ ($V_0=0.14t,\, \epsilon_d=E_F-0.5t$, $U=1.0t$, and $t=0.25$) and $8$ (a) and $7$ (b)
electrons. Lower panels are the NRG results with $\alpha =0$, $N_{c}=13$ (c)
and $N_c=14$ (d). ($\varepsilon _{d}=-0.5$, $U=1.0$, and $V=0.2$.)}
\label{fig5}
\end{figure}

Using a Lanczos algorithm we first calculate the ground state energy and
wavevector. In the present case it is convenient to use a second Lanczos
algorithm\cite{Gagliano89} to evaluate directly the impurity spectral function, rather than
evaluating excited states, their corresponding energies and matrix
elements $M_{\lambda ,\lambda ^{\prime }}$. For a better comparison of the
different cases, we always measure the frequency from the Fermi energy $E_{F}
$ defined as $E_{F}=[E_{0}(N_{e}+1)-E_{0}(N_{e}-1)]/2$ with $E_{0}(N_{e})$
the ground state energy of the system with $N_{e}$ particles. The spectral
density around the Fermi energy shows a series of peaks separated by the
characteristic energy $\Delta $. Each peak may be composed by a one or a few
delta lines as shown in Figs. \ref{fig5}(a) and \ref{fig5}(b). In what follows we discuss the
low energy structure ($\omega \lesssim \Delta $) of the impurity spectral
function for the case of even and odd number of particles corresponding to
the at-resonance and off-resonance situations respectively. Note that the results of Figs. \ref{fig5}(a) and \ref{fig5}(b) are not for a half filed system and as a consequence the electron-hole symmetry is not exactly preserved.

{\it Even number of particles:} The ground state of the system is a singlet
and the impurity spectral density shows a central peak at $\omega \sim 0$
and two satellites at $\omega \sim \pm \Delta $. The central peak consists
of two lines, one for $\omega >0$ and one for $\omega <0$ corresponding to
adding and subtracting a particle respectively. The splitting between these
two lines, that decreases as $V_{0}$ decreases, is not given directly by the
confinement energy $\Delta $ but by the energy gained by forming the
singlet, i.e. is given by the Kondo temperature of the finite system.
The peaks at $\omega \sim \pm \Delta $ are also made of a couple of delta
lines each. The justification of this description of a central line and
satellites at $\pm \Delta $ is given by the fact that for smaller
hybridizations the two central lines approach $\omega \sim 0$ and the
impurity spectral density reproduces the structure of the unperturbed local
density of states (the at-resonance situation of Fig. \ref{fig2}(a)). Here we
used these parameters for a better comparison; results with a
smaller hybridization calculated with the NRG are shown below.

{\it Odd number of particles: }The ground state of the system is a spin-1/2
doublet. The spectral densities have a low frequency gap with peaks at $%
\omega \sim \pm \Delta /2$ . Each of these peaks at $\omega \sim \pm \Delta
/2$ consist of two delta lines that correspond to final states with
different total spin: when an electron is created or destroyed on the
spin-1/2 ground state, the final state may have total spin zero or one. The
energy of the final state depends on its total spin. Again in this case the
general structure is that of the underlying local density of states.

These results should serve us as a guide for the numerical renormalization
group calculation. Since the NRG involves some approximations and will be
extended to more realistic cases, as is the case of a system with a finite
barrier, it is important to have this exact result as a reference
calculation.

\subsection{Numerical Renormalization Group: Zero Temperature Results}

For a finite system, the renormalization procedure is truncated at an
iteration $N_{c}$. In Figs. \ref{fig5}(c) and \ref{fig5}(d) we present the NRG results for
the impurity spectral density of a finite system with an even and an odd
number of particles respectively. As for the calculation of the
thermodynamic properties, the system with an even (odd) number of particles
is evaluated with an odd (even) number of shells $N_{c}$. The spectrum
consists of a discrete collection of delta lines and, as in the previous
section, the smoothing is done only for practical purposes in the
presentation of the data. Our low frequency NRG results compare very well
with the exact results of a finite system [Figs. \ref{fig5}(a) and \ref{fig5}(b)]. Again, for
an even number of particles the impurity spectral density consists of a
central peak, composed by two delta functions, and the two satellites at $%
\omega \sim \pm \Delta $. For an odd number of particles the approximate NRG
results clearly show the central pseudogap and the two satellites at $%
\omega \sim \pm \Delta /2$ with a structure that comes from the
spin-dependent energy of the final state.

For the remainder, we focus on the more interesting case of a system in
contact with a macroscopic reservoir. This is described with a non-zero
parameter $\alpha $ representing a finite wall.

\begin{figure}[tbp]
\includegraphics[width=7cm,clip=true]{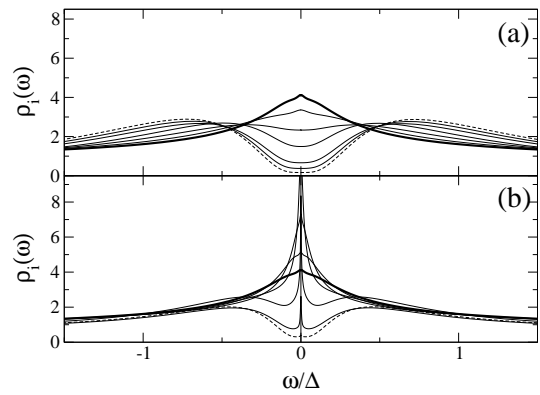}
\caption{Impurity spectral density $\rho _{i}(\omega ,T=0)$ for the at-resonance case with $N_{c}=19$ (a) and the off-resonance case with $N_{c}=20$
(b) and different values of $\alpha $: $\alpha =0.2$ (dashed line), $0.3$, $0.4$ , $0.6$, $0.75$, and $0.9$ (thick line). ($\varepsilon _{d}=-0.5$, $U=1.0$, and $V=0.2$.)}
\label{fig6}
\end{figure}

In Fig. \ref{fig6}(a) the impurity spectral density for the at-resonance situation is
shown for different values of $\alpha $. For $\alpha =0$ the isolated cluster
results are reproduced with a peak at low frequencies. This peak is well
separated from the Fermi energy due to a large hybridization $V$ used in the
calculation for practical purposes. In the figure only a detail of the low
frequency structure is shown and the satellite at $\omega \sim \Delta $ is
not observed. As $\alpha $
approaches one, corresponding to an infinite homogeneous system, the two
structures merge into a single Kondo peak centered at the Fermi energy. 
The impurity spectral density when the Fermi energy is off-resonant is shown
in Fig. \ref{fig6}(b). For a small $\alpha$ the excitations around $\omega \sim \Delta /2$ are clearly observed. As $\alpha $ increases the
$\omega \sim \Delta /2$ structure is washed out and simultaneously a narrow Kondo
resonance develops at the Fermi energy. 

\begin{figure}[tbp]
\includegraphics[width=7cm,clip=true]{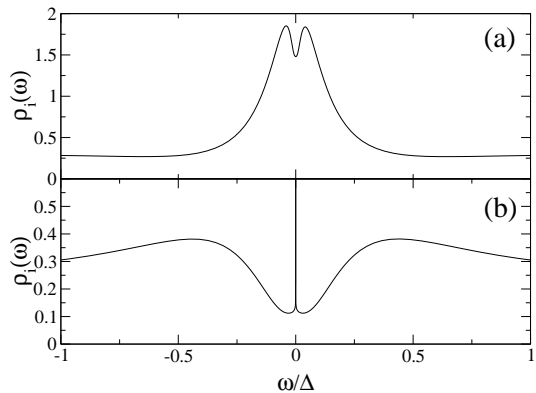}
\caption{ Same as in Fig. \ref{fig6} with $\alpha =0.45$ and $V=0.15$. a) At-resonance
case with $N_{c}=19$ and b) Off-resonance case with $N_{c}=20$.}
\label{fig7}
\end{figure}

For more realistic parameters (a smaller hybridization and consequently a
smaller Kondo temperature) the at-resonance and off-resonance impurity
spectral densities are shown in Fig. \ref{fig7}. These results show that, for the
general case of a Kondo impurity in a nanoscopic system with an intermediate
barrier and the Fermi energy at a resonant state, we should expect the low
frequency spectral density $\rho _{i}(\omega ,T=0)$ to present a broad
Kondo resonance with some structure. In fact, for large barriers (small $\alpha$), the Kondo resonance has a minimum at the Fermi energy. For the off-resonant case, we expect the spectral density to
present the structures at $\omega \sim \pm \Delta /2$ separated by a
pseudogap and - at zero temperature - a central Kondo peak. 
As we show in the next section, a small temperature completely destroys this
central Kondo peak while the broad structures survive up to much higher
temperatures.
\subsection{Phenomenological approach}

\begin{figure}[tbp]
\includegraphics[width=7cm,clip=true]{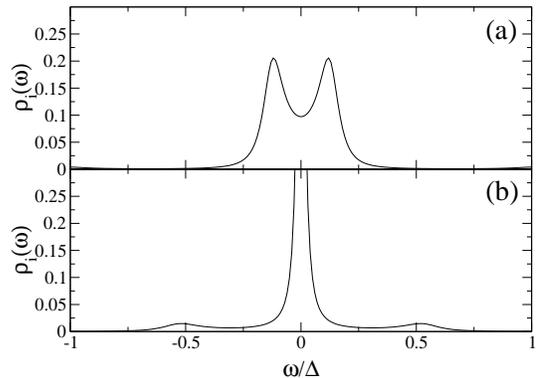}
\caption{ Phenomenological results for the impurity spectral densities. ($\Delta=1.0$, $\gamma=0.1$, and $b^2V^2=0.02$). a) At-resonance. b) Off-resonance. }
\label{fig8}
\end{figure}

Here we briefly discuss the structure of the Kondo resonance in terms of what would be obtained with a simple
slave boson theory in the saddle point approximation. In this theory ($U=\infty$), the low energy impurity Green's function is given by \cite{Hewson1993}
\begin{equation}
G \left( \omega \right) = \frac{b^2}{\omega -\varepsilon_\lambda - b^2V^2g^0\left(\omega\right)}
\end{equation}
where $b^{2}$ is the square of the mean value of the boson field, $\varepsilon _{\lambda }\simeq 0$ gives the position of the Kondo resonance and $g^{0}(\omega )$ is the bare local propagator of conduction electrons.
Here as in the NRG calculation we take the Fermi energy equal to zero. In an homogeneous system it is a good approximation to take $g^{0}(\omega )=-i\pi \rho$
where $\rho $ is the frequency independent local density of
states and defining the Kondo temperature $T_{K}$ as the width of the Kondo resonance the local propagator can be
put as $G(\omega )\simeq b^{2}/(\omega +iT_{K})$ with $b^{2}=T_{K}/\pi V^{2}\rho $ in agreement with the Friedel's sum
rule. For our case of a central grain weakly coupled with a reservoir, the propagator $g^{0}(\omega)$ is taken
as a sum of poles separated by the characteristic energy $\Delta$ and widths $\gamma$
\begin{equation}
g^{0}(\omega
)=\sum_{l}\frac{1}{\omega -\Delta _{l}+i\gamma }
\end{equation}
with $\Delta _{l}=l\Delta $ or $\Delta
_{l}=(l+1/2)\Delta $ for the at-resonance and off-resonance cases respectively. 
The structure of the Kondo
resonance so obtained is shown in Fig \ref{fig8}. 
 The results are in good qualitative agreement with the NRG results. In the at-resonance case, depending on the value of the parameters, the low energy spectrum consists of two peaks with a deep at the Fermi level or a single peak with a maximum at the Fermi level (not shown). In the off-resonance case, a central peak at the Fermi level is obtained.  Note however that the area of the central peak relative to the area of the satellite peaks obtained is this approximation is different from that obtained with the NRG.
\subsection{Finite Temperature Results}

The NRG calculation of the finite temperature spectral density $\rho_{i}(\omega,T)$ relies on the same approximations of the $T=0$ case. The spectral density at a fixed temperature $T$ is evaluated as above - using Eq. (\ref{ai}) instead of Eq. (\ref{ai0}) - if $\omega >k_BT$. To calculate the spectral density at $\omega \lesssim k_{B}T$ a Hamiltonian $H_{N}$ with $N$ such that $\tilde{\omega} _{N}\sim k_{B}T$ is used.

When $T_K^\infty \sim \Delta$ and the Fermi energy lies between resonances, the off-resonance case, two energy scales are clearly observed in the
temperature dependence of the susceptibility. In fact as the temperature is
lowered they determine the onset of the plateau in $k_{B}T\chi$, that is given by $\Delta$, and the offset at $T_K^{eff}$ that is determined by $\alpha$.
 For the at-resonance case the static magnetic properties are
dominated by the larger scale $\Delta $ since for $k_{B}T<\Delta $ the
magnetic moment is completely screened. The spectral densities are also
sensitive to the temperature and as we will see, there are changes in $\rho
_{i}(\omega ,T)$ each time the temperature approaches each one of the
characteristic energy scales.

\begin{figure}[tbp]
\includegraphics[width=7cm,clip=true]{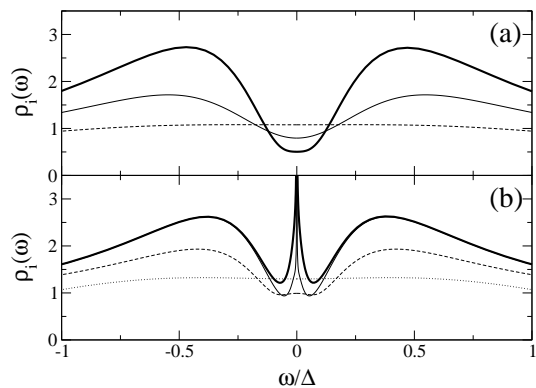}
\caption{Impurity spectral density $\rho _{i}(\omega ,T)$ for different temperatures. (a) At-resonance: $T=0$ (thick line), $T=0.23\Delta$ (thin line), and $T=0.7\Delta$ (dashed line) (b) Off-resonance: $T=0.0$ (thick line), $T=2.0\times 10^{-3}\Delta$ (thin line), $T=5.1\times 10^{-2}\Delta$ (dashed line), and $T=0.5\Delta$ (dotted line). }
\label{fig9}
\end{figure}

In Fig. \ref{fig9}(a) the spectral density $\rho_{i}(\omega ,T)$ is shown at different
temperatures for a system with the parameters as in Fig. \ref{fig6}.
The Kondo peak develops and reaches its zero temperature value as soon as the temperature goes below $\Delta$. The parameters are the same as in Fig. \ref{fig3} with $\alpha=0.35$.
 
For the off-resonant situation the temperature evolution is shown in
Fig. \ref{fig9}(b). In this last case the two characteristic temperatures are also reflected in the magnetic susceptibility from which we can extract their
values as the onset and offset of the plateau in $\mu ^{2}$. As can be seen
from the figure, as the temperature is lowered, the broad satellite at $%
\omega \sim \Delta /2$ develops at a temperature corresponding to the onset
of the plateau while a narrow Kondo structure develops at a temperature
corresponding to the offset of the plateau. The first and second temperatures at which the narrow Kondo peak and the broad structure disappear indicate the decoupling of the impurity spin with the electron spin density at large ($r>R_{c}$) and short ($r<R_{c}$) distances respectively.
\section{Summary and Discussion}
We have presented a model for a Kondo impurity in a nanoscopic sample or
grain coupled with a reservoir through a large barrier. The natural energy -
or length - scale associated to the Kondo effect in a bulk material is to be
compared with the new scale introduced by the confinement of electrons.
These new scales modify the local density of states at the impurity site
and consequently their thermodynamic and spectral properties. 

The local density of states of the host material consists of resonant states
separated by a characteristic energy $\Delta $ and with a width $\gamma $.
While $\Delta $ is determined by the size ($R_{c}$) of the grain, the width
of the resonances is determined by the transparency ($\alpha $) of the
barrier. We have considered cases in which the Fermi energy lies at a
resonance or between two of them. 

The main results of the paper are the frequency and temperature dependence
of the impurity spectral density $\rho _{i}(\omega ,T)$. We have performed a
detailed analysis of the structure of $\rho _{i}(\omega ,T)$ for $\omega $
around the Fermi energy for the case $T^\infty_K\sim \Delta$. This regime with $T^\infty_K\sim \Delta$ is characterized by a interplay between electron-electron correlation and confinement effects. The width of the Kondo resonance ($T^\infty_K$) and the characteristic energy ($\Delta$) that defines the structure of the local density of states are of the same order and as a consequence the Kondo resonance shows a superstructure. When the Fermi energy lies at a resonance, the low temperature spectral density consists of a broad structure around the Fermi level. For large barriers, the spectrum presents a two peaked structure with a deep at the Fermi level, where the distance between peaks is given by the Kondo energy of the finite system. As the height of the barrier is decreased, the peaks merge into a single broader peak.

For the off-resonance case $\rho _{i}(\omega ,T=0)$ presents structures at $%
\omega \sim \pm \Delta /2$ and a central Kondo-like peak at $\omega \sim 0$.
In temperature dependence two energy scales can be distinguished, at which the two types of structures disappear. These two energy scales $\Delta$ and $T_K^{eff}$ are also reflected in the magnetic susceptibility. 
Phenomenological results based on the slave boson mean field theory for the impurity spectrum at zero temperature give results in qualitative agreement with the ones obtained with NRG.

Different experiments like the STM conductance for impurities in quantum
corrals or small terraces or the transport through QD in nanoscopic rings
could be used to test the theory.

This work was partially supported by the CONICET and ANPCYT,
Grants N. 02151 and 99 3-6343.


\begin{thebibliography}{33}
\expandafter\ifx\csname natexlab\endcsname\relax\def\natexlab#1{#1}\fi
\expandafter\ifx\csname bibnamefont\endcsname\relax
  \def\bibnamefont#1{#1}\fi
\expandafter\ifx\csname bibfnamefont\endcsname\relax
  \def\bibfnamefont#1{#1}\fi
\expandafter\ifx\csname citenamefont\endcsname\relax
  \def\citenamefont#1{#1}\fi
\expandafter\ifx\csname url\endcsname\relax
  \def\url#1{\texttt{#1}}\fi
\expandafter\ifx\csname urlprefix\endcsname\relax\def\urlprefix{URL }\fi
\providecommand{\bibinfo}[2]{#2}
\providecommand{\eprint}[2][]{\url{#2}}

\bibitem{nanoscience}{D. Goldhaber-Gordon, H. Shtrikman, D. Mahalu, D.
Abusch-Magder, U. Meirav, and M.A. Kaster, Nature (London) {\bf 391}, 156
(1998); C. Dekker, Phys. Today {\bf 52}, 22 (1999).}

\bibitem{Hewson1993}  For a review, see A.C. Hewson, {\it The Kondo problem to
heavy fermions} (Cambridge University Press, Cambridge, England, 1993).


\bibitem{madhavan98}  V. Madhavan, W. Chen, T. Jamneala, M.F. Crommie, and N.S.
Wiegreen, Science {\bf 280}, 567 (1998).

\bibitem{nagaoka2002}  K. Nagaoka, T. Jamneala, M. Grobis, and M.F. Crommie, Phys. Rev. Lett. {\bf 88}, 077205 (2002).

\bibitem{Manoharan2000}  H.C. Manoharan, C.P. Lutz, and D.M. Eigler, Nature {\bf 403} 512 (2000).

\bibitem{Meir1991}  Y. Meir, N.S. Wingreen, and P.A. Lee, Phys. Rev. Lett. {\bf 66}, 3048 (1991).

\bibitem{Wilson1975}  K.G. Wilson, Rev. Mod. Phys {\bf 47}, 773 (1975); H.R.
Krishna-murthy, J.W. Wilkins and K.G. Wilson, Phys. Rev. B {\bf 21}, 1044
(1980).

\bibitem{Thimm1999}  W.B. Thimm, J. Kroha, and J. von Delft, Phys. Rev. Lett. {\bf 82}, 2143 (1999).

\bibitem{Cornaglia2002} P.S. Cornaglia and C.A. Balseiro, cond-mat/0202489, Phys. Rev. B (to be published). 

\bibitem{Affleck2001}{I. Affleck and P. Simon, Phys. Rev. Lett. \textbf{86}, 2854 (2001); Hiu Hu, Guang-Ming Zhang, and Lu Yu, Phys. Rev. Lett. \textbf{86}, 5558 (2001); Ian Affleck, cond-mat/0111321.}

\bibitem{Anderson1961}  P.W. Anderson, Phys. Rev. {\bf 124} 41 (1961).

\bibitem{Nozieres} {P. Nozi\`eres, {\it Proceedings of the 14th. International Conference on Low Temperature Physics}, (Ed. M. Krusius and M. Vuorio, North-Holland, Amsterdam, 1975), Vol. 5, p 339}

\bibitem{Desc} {J. des Cloizeaux, in {\it Theory of Consensed Matter}, (IAEA, Vienna, 1968). }

\bibitem{Hofstetter} {T. A. Costi, A. C. Hewson and V. Zlatic, J. Phys. Cond. Matt.{\bf 6}, 2519 (1994); Walter Hofstetter, {\it Renormalization Group Methods for Quantum Impurity Systems}, (Shaker Verlag, Aachen, Germany, 2000).}

\bibitem{Gagliano89} E.R. Gagliano and C.A. Balseiro, Phys. Rev. Lett. {\bf 59}, 2999 (1987).


\end{thebibliography}
\end{document}